# Passive Daytime Radiative Cooling Enabled by Bio-Derived Ceramic–Polymer Coatings on Rapid-Curing Fiberglass Casts

Xuguang Zhang[a], Hexiang Zhang[a], Hanqing Liu[a], Xiaoli Li[a], Mu Ying[a], Yutian Yang[b], Marilyn L. Minus[a], Ming Su[b], Yi Zheng[a,b,*]

[a] Department of Mechanical and Industrial Engineering, Northeastern University, Boston, MA 02115, USA

[b] Department of Chemical Engineering, Northeastern University, Boston, MA 02115, USA

[*] Corresponding author: y.zheng@northeastern.edu

## ABSTRACT

Passive daytime radiative cooling (PDRC) provides an energy-free approach to suppress surface temperatures by reflecting solar irradiation while emitting thermal radiation through the mid-infrared atmospheric window. Despite rapid progress in optical performance, most PDRC systems remain limited by rigid, fragile, or planar substrates, restricting their use on flexible, curved, or wearable surfaces. Here, we report a biocompatible and structurally robust PDRC system integrated onto a commercial rapid-curing fiberglass cast, a conformal substrate widely used in orthopedic and industrial applications. The cooling architecture adopts a bilayer polymer design consisting of a polyvinyl alcohol (PVA) adhesion layer and a polymethyl methacrylate (PMMA) protective layer, both embedded with calcium pyrophosphate (CPP) ceramic particles derived from processed animal bone waste. The bio-derived CPP simultaneously enables broadband solar scattering and high mid-infrared emittance, while offering sustainability and biocompatibility advantages. The resulting composite exhibits over 90% solar reflectance and achieves up to 15 °C sub-ambient cooling under direct outdoor sunlight. Beyond optical performance, the coated fiberglass cast demonstrates strong environmental and mechanical durability, including enhanced water resistance (contact angle ≈ 85°), ultraviolet stability, abrasion tolerance, and thermal stability exceeding 700 °C. Tensile testing further confirms that the bilayer coating improves flexibility while preserving structural integrity, allowing conformal deployment on irregular and anatomically curved surfaces. By combining photonic thermal control with bio-derived materials and a rapid-curing, flexible substrate, this work establishes a scalable pathway for wearable thermal regulation, orthopedic comfort enhancement, and mobile passive cooling in medical and outdoor environments.

1. Introduction

The rapid expansion of urbanization and the intensification of global warming have driven a critical need for sustainable, passive cooling strategies that can mitigate rising surface temperatures without relying on electricity or active ventilation [1-4]. Among these, passive daytime radiative cooling (PDRC) has emerged as a particularly promising approach, leveraging the Earth's natural heat balance to reject solar radiation and emit thermal energy through the atmospheric transparency window (8–13 μm) [5–8]. By achieving both high solar reflectance and strong mid-infrared emittance, PDRC materials can maintain sub-ambient temperatures even under direct sunlight, making them ideal for energy-saving building envelopes, thermal shielding of infrastructure, and mobile off-grid systems [9–11].

Despite substantial advances in laboratory-scale PDRC materials, ranging from hierarchical polymers to porous ceramics, translating these technologies into practical outdoor use remains challenging [12-14]. Many reported systems require rigid substrates, multi-step fabrication, or delicate nano-structuring, which limit their flexibility, scalability, or mechanical integrity under real-world conditions [15–16]. Moreover, most coatings degrade rapidly in harsh environments due to UV exposure, water infiltration, or mechanical wear [17-18]. These limitations underscore the pressing demand for PDRC systems that are not only optically efficient but also physically robust, weather-resistant, and compatible with non-planar or deformable surfaces.

In this work, we present a structurally adaptive, multifunctional PDRC system constructed on a commercial fiberglass cast, a material originally designed for orthopedic and industrial reinforcement, but here innovatively repurposed as a flexible and field-deployable substrate.

The fiberglass cast comprises a woven fiberglass matrix pre-impregnated with a hydrated ester resin, enabling rapid curing under both air and moisture exposure. Such features allow conformal integration onto irregular surfaces, while providing a strong structural backbone for the cooling layers. Compared to conventional rigid or flat PDRC substrates, the use of fiberglass cast enables seamless deployment in curved, mobile, or wearable settings.

Beyond structural compatibility, the integration of PDRC into fiberglass casts opens new opportunities in healthcare applications. Fiberglass casts are commonly used in orthopedics to immobilize injured limbs; however, prolonged use in hot climates or high-humidity environments can lead to thermal discomfort, sweating, or skin irritation. Incorporating radiative cooling functionality directly into the cast material provides passive temperature regulation at the skin–device interface, potentially enhancing patient comfort, preventing heat-induced inflammation, and reducing the risk of dermatological complications during extended wear. This direction represents a previously unexplored intersection between biomedical support devices and thermal photonic engineering.

Moreover, the biocompatible, lightweight, scalable, and water-curable nature of the proposed system makes it a strong candidate for broader thermal management applications in resource-limited settings [19]. These include emergency shelters in disaster relief zones, portable enclosures for medical and food storage, and wearable thermal protection for outdoor laborers or first responders [20-21]. The versatility of the coating architecture, combined with its demonstrated thermal, mechanical, and environmental resilience, suggests that PDRC-integrated fiberglass systems can contribute meaningfully to the next generation of adaptive, infrastructure-independent cooling technologies [22-24].

## 2. Fabrication and Characterization.

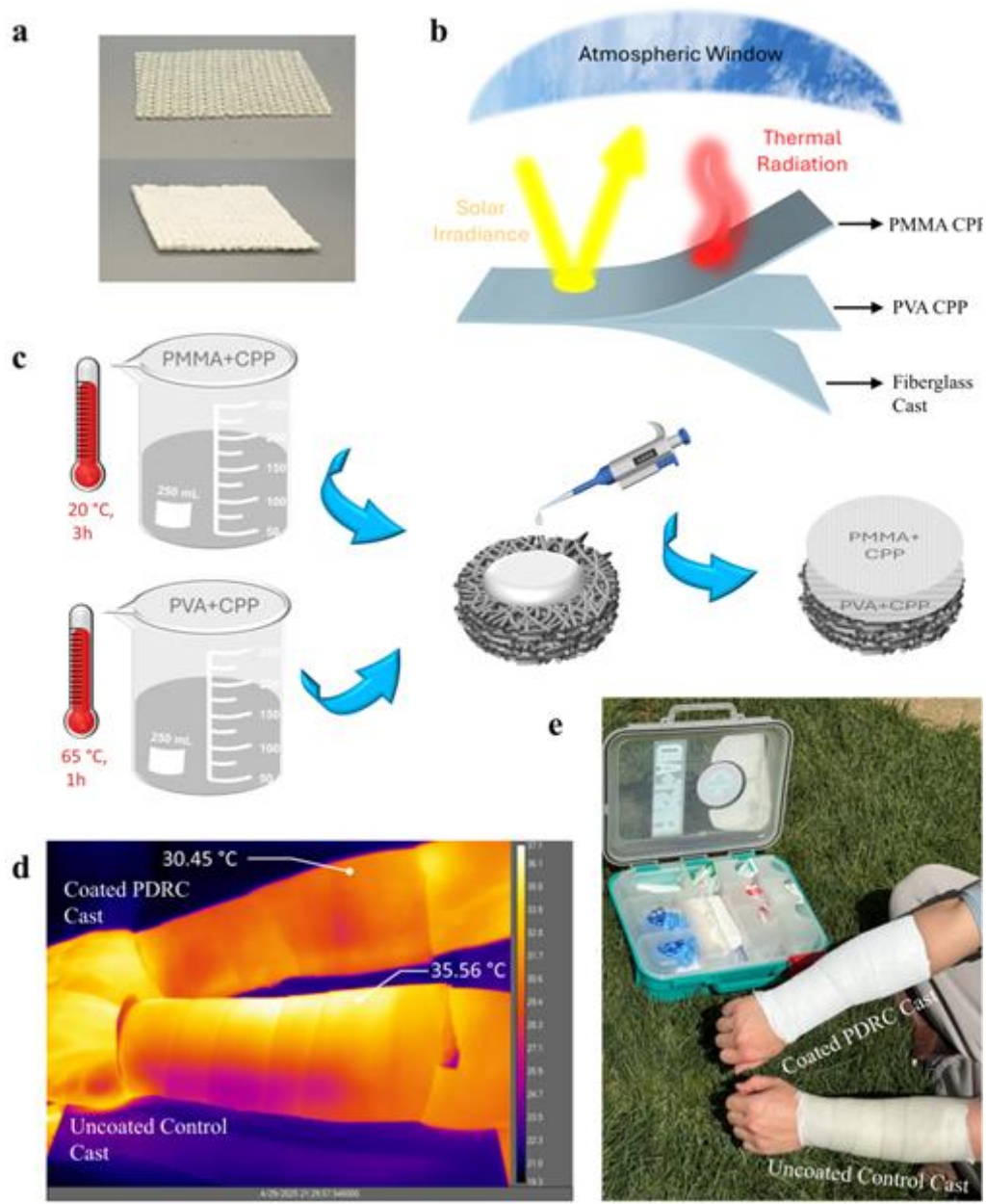

**Figure 1.** Fabrication and application of the bilayer PDRC coating on fiberglass cast. (a) Commercial fiberglass cast substrate composed of woven fiberglass and hydrated ester resin. (b) Schematic of the bilayer coating: PVA base for adhesion, PMMA protective layer for water resistance, both embedded with CPP particles. (c) Fabrication process: solution mixing, coating, drying, and curing. (d) Infrared image under solar simulator showing lower steady-state temperature 30.45 °C for the coated sample versus 35.56 °C for the uncoated cast. (e) Outdoor demonstration of the coated cast worn on a human subject under sunlight.

The substrate employed in this study is a commercial fiberglass cast composed of woven fiberglass reinforced with a hydrated ester resin system. This material exhibits rapid solidification behavior under both ambient air and water exposure. When immersed in water, the cast cures within several minutes, while under ambient air, full solidification typically occurs within 30–60 minutes. These properties make it suitable for structural integration in flexible, field-deployable passive cooling systems.

As shown in Figure 1b, a bilayer PDRC structure was developed on the surface of the fiberglass cast. The bottom layer is a PVA-based coating designed to offer high extensibility and strong conformability, ensuring intimate adhesion to the curved or irregular surface of the fiberglass cast. The protective layer is based on PMMA, selected for its hydrophobicity and environmental resistance, providing surface protection and stability under outdoor conditions.

To fabricate the PVA-based bottom layer, PVA powder was dissolved in deionized water at a 1:10 mass ratio. The solution was stirred magnetically at 700 rpm and heated at 65 °C for 1 hour until it became transparent. Bio-derived CPP particles, obtained from thermally treated animal bone waste, were then added to serve as the solar-reflective pigment [25]. After 15 minutes of additional stirring, the PVA-CPP mixture was uniformly applied to the outer surface of the fiberglass cast using a pipette. Owing to the low viscosity of the solution, the coating demonstrated self-leveling behavior, forming a continuous film without the need for additional processing. The layer was allowed to dry under ambient conditions for approximately 8 hours.

Once the bottom layer was fully dried, the PMMA protective layer was prepared by dissolving PMMA in acetone at a 1:10 mass ratio. The solution was stirred for 3 hours until transparent, followed by the addition of CPP particles and an additional 15 minutes of mixing. Due to the rapid evaporation of acetone, the PMMA-CPP solution exhibited high viscosity and fast solidification. The mixture was applied to the dried PVA layer using a brush to ensure uniform surface coverage. This protective coating solidified in ambient air within approximately 1 hour. The full fabrication process is summarized schematically in Figure 1c.

To further validate the cooling effectiveness of the bilayer-coated fiberglass cast in a realistic usage scenario, thermal imaging experiments were conducted under both controlled and natural conditions. As shown in Figure 1d, under a solar simulator, the surface temperature of the coated fiberglass cast worn on the human forearm stabilized at 30.45 °C, significantly lower than the 35.56 °C observed on the uncoated fiberglass under identical conditions. This temperature difference of over 5 °C confirms the efficacy of the radiative coating in reducing surface heat buildup under simulated sunlight. Complementary results were observed in Figure 1e, which presents an outdoor application scene where a volunteer wears the cast in direct sunlight. The visibly brighter infrared signal from the uncoated region further supports the coating's ability to suppress heat accumulation on the skin–device interface, highlighting its potential to enhance wearer comfort during prolonged use.

To evaluate the structural integrity and functional performance of the bilayer PDRC-fiberglass cast composite, a comprehensive set of characterizations was conducted. The spectral properties were measured across two key wavelength regions: the ultraviolet–visible–near-infrared (UV–Vis–NIR) range of 0.2–2.5 μm and the mid-infrared (MIR) range of 2.5–18 μm. These measurements provide insights into the material's solar reflectance and thermal emittance, which are critical for radiative cooling applications. Real-world thermal performance was validated through outdoor field tests under direct sunlight. To assess long-term environmental durability, the composite underwent ultraviolet (UV) exposure and mechanical wear tests. Mechanical robustness was further examined through two separate evaluations: mechanical resistance and mechanical stretch testing, aimed at simulating deformation and abrasion scenarios that the coating may encounter during use. Surface wettability was evaluated using static water contact angle measurements to assess the hydrophobicity of the PMMA protective layer. In addition, thermogravimetric analysis (TGA) was conducted to examine the thermal stability of the bilayer system. Porosity was analyzed to investigate microstructural features that may influence optical scattering and filler dispersion. These combined tests confirm that the fabricated bilayer structure possesses the fundamental attributes required for passive radiative cooling, including spectral selectivity, mechanical resilience, and environmental robustness.

## 3. Results and discussion

### 3.1. Indoor Optical and Thermal Radiative Test

To evaluate the radiative cooling performance of the PDRC-fiberglass system under controlled conditions, a series of indoor spectral measurements were conducted. These tests served two purposes: first, to determine the optimal concentration of CPP as a radiative pigment; and second, to compare the spectral performance of CPP with other widely used PDRC fillers.

Figure 2a shows the spectral reflectance of the fiberglass cast coated with only the pure binding matrix materials (PVA and PMMA). The addition of these polymers alone does not significantly alter the spectral performance of the substrate, and the reflectance remains low across the solar spectrum (0.3–2.5 μm), consistent with the baseline behavior shown in Figure 2c. However, once CPP is introduced as a filler, the reflectance markedly increases with increasing concentration. As shown in Figure 2b, CPP concentrations ranging from 5 wt% to 30 wt% were tested on bilayer (PVA + PMMA) coated fiberglass casts. A clear trend is observed: reflectance improves steadily as CPP loading increases, with the 30 wt% sample achieving over 90% average reflectance in the solar spectrum. Based on this balance between optical performance and material cost, 30 wt% CPP was selected as the optimal formulation for subsequent tests.

After optimizing the CPP concentration, a comparative study was performed using other common PDRC fillers, including titanium dioxide ($TiO_2$), zinc oxide (ZnO), zirconium dioxide ($ZrO_2$), hydroxyapatite (HAP), and polytetrafluoroethylene (PTFE). As shown in Figure 2c, the original uncoated fiberglass cast exhibited poor performance, with reflectance around 30%. All other samples, coated with bilayer polymer matrices and loaded with 30 wt% of each filler, show improved reflectance across the 0.3–2.5 μm range. Among them, CPP demonstrates the highest reflectance, followed closely by $TiO_2$. The remaining materials (HAP, PTFE, ZnO, $ZrO_2$) exhibit comparable but moderately lower reflectance levels.

In addition to the solar spectral region, the mid-infrared (MIR) range from 8–13 μm was examined, which corresponds to the second atmospheric transparency window crucial for radiative heat dissipation to outer space. The reflectance data in this range are also shown in Figure 2c. As radiative cooling performance in this band is governed by Kirchhoff's law of thermal radiation:

$$\varepsilon(\lambda) = 1 - R(\lambda) \quad (1)$$

where $\varepsilon(\lambda)$ is the spectral emittance and $R(\lambda)$ is the spectral reflectance at wavelength $\lambda$, the low reflectance observed in this range across all samples implies high emittance. This indicates that the tested coatings are capable of effectively radiating thermal energy through the MIR atmospheric window.

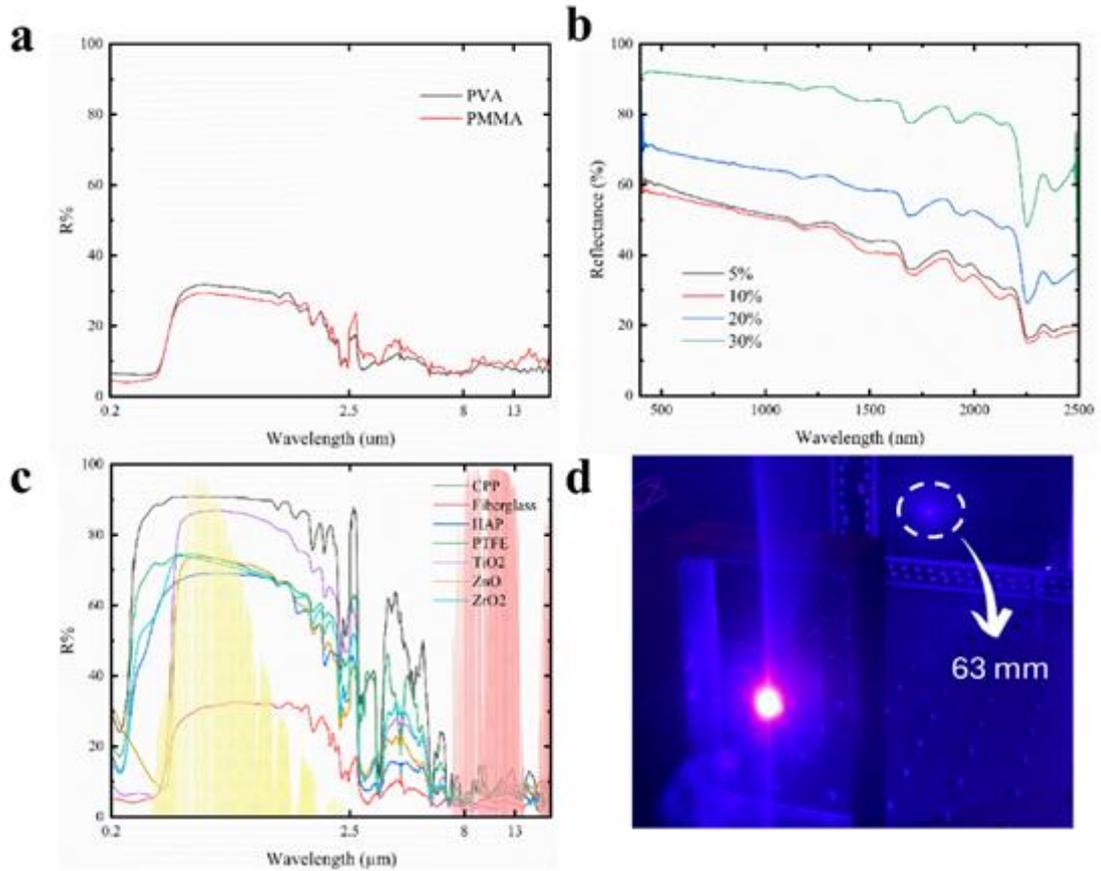

**Figure 2.** Indoor spectral characterization and CPP optimization. (a) Solar reflectance of fiberglass casts coated with pure PVA and PMMA layers, showing minimal enhancement compared to uncoated substrate. (b) Solar reflectance of bilayer coatings with increasing CPP concentrations (5–30 wt%), revealing performance improvement with higher filler content. (c) Comparison of solar (0.3–2.5 μm) and mid-infrared (8–13 μm) reflectance for different fillers (CPP, TiO$_2$, ZnO, PTFE, ZrO$_2$, HAP) on fiberglass cast; CPP yields the highest solar reflectance and strong mid-IR emittance. (d) UV laser scattering test comparing CPP powder, tape, fiberglass cast, and air. The CPP-coated sample produces a significantly larger scattering spot, confirming strong short-wavelength scattering capability.

Furthermore, Figure 2d presents a UV laser scattering experiment conducted to qualitatively assess the optical scattering behavior of CPP powder. A thin layer of CPP was adhered to transparent tape and irradiated with a UV laser source positioned 20 cm from the sample. The resulting spot was projected onto a screen placed 30 cm behind the sample. The scattering diameter on the screen was 63 mm for the CPP-coated tape, compared to 36 mm for pure tape, and only 20 mm for the fiberglass cast sample. For reference, the original laser beam had a spot diameter of 5 mm. These results confirm that CPP has a strong scattering capability for short-wavelength radiation, contributing to its high solar reflectance.

*3.2. Outdoor Passive Radiative Cooling Test*

To evaluate the real-world PDRC performance of the bilayer-coated fiberglass samples, two sets of outdoor temperature experiments were conducted under different environmental conditions. These tests serve to verify the effectiveness of the materials in reducing surface temperature under direct solar irradiation, as well as their applicability in practical outdoor scenarios.

The first test was conducted over a continuous 62-hour period from the afternoon of October 15, 2024, to the evening of October 18, 2024. The experiment took place at the College of Engineering, Northeastern University (42.3384° N, 71.0890° W), Boston, MA. As shown in Figure 3a and 3b, the samples were mounted on an aluminum foil base to ensure thermal insulation from below. Small perforations were made in the aluminum layer to allow thermocouples to directly contact the underside of each sample, ensuring accurate temperature acquisition. The entire setup was enclosed within an acrylic chamber to minimize the influence of wind, which can introduce convective cooling and mask the true radiative cooling effects of the materials. This wind isolation approach allows the test to more accurately reflect the cooling potential of the materials in a still-air environment, which mimics many real-world applications such as rooftop coatings, tents, or temporary shelters.

The tested samples included fiberglass casts coated with 30 wt% of various PDRC fillers (CPP, $TiO_2$, ZnO, PTFE, $ZrO_2$, and HAP) following the bilayer structure described earlier. In addition, two reference samples, bare wood and commercially painted white wood, were used as practical baselines. The ambient air temperature was recorded simultaneously and used as a reference for cooling performance comparison. Solar irradiance data, also shown in Figure 3c, was collected via a nearby weather station and plotted alongside the sample temperatures.

During daytime peak conditions, the CPP-coated sample consistently maintained the lowest temperature among all test samples. On the first full day, when the ambient temperature peaked at approximately 55 °C, the CPP sample remained at 42 °C, showing a significant cooling effect of 13 °C. $TiO_2$ and painted wood also exhibited cooling, but their peak temperatures were still approximately 2 °C higher than that of CPP. On the second day, a similar trend was observed. The maximum temperature difference between CPP and the ambient occurred at hour 55, with the ambient reaching 57 °C while the CPP sample remained at 42 °C, yielding a 15 °C cooling differential.

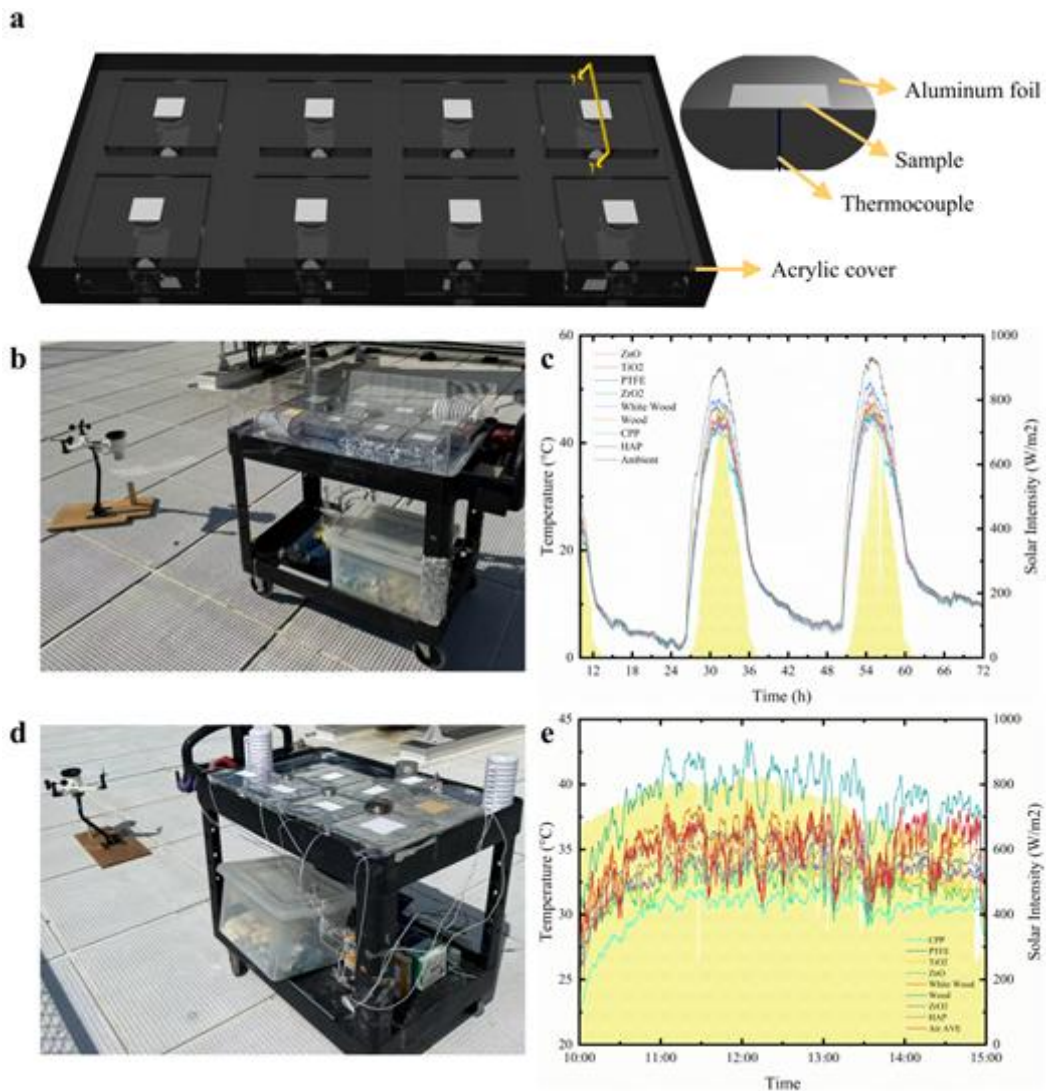

**Figure 3.** Outdoor temperature tests under controlled and natural conditions. (a) Schematic of the covered outdoor test setup with samples mounted on aluminum foil and thermocouples inserted below each sample. (b) Experimental photograph of the covered setup with acrylic enclosure to minimize wind effects. (c) Temperature tracking over 62 hours showing the CPP-coated sample maintains up to 15 °C sub-ambient cooling; ambient temperature peaks at 57 °C. (d) Setup of the uncovered outdoor test conducted under natural sunlight and wind exposure. (e) Temperature profiles from the uncovered test showing CPP again achieves the lowest surface temperature (~5 °C below ambient) despite environmental fluctuations.

Another group of the outdoor test was performed without wind shielding to further assess material performance under more realistic and variable environmental conditions. This experiment was conducted during the daytime of August 13, 2024, under clear and sunny weather. The experimental configuration is shown in Figure 3d, and the recorded data are presented in Figure 3e. The same set of samples was used. Compared to the shielded setup, the unshielded test exhibited larger temperature fluctuations due to the influence of ambient wind. Nonetheless, the CPP-coated sample continued to demonstrate the best cooling performance. During the peak period, the ambient temperature reached 38 °C, while the CPP surface stabilized at 32 °C, maintaining a 6 °C temperature difference. Other materials exhibited weaker performance, with larger fluctuations and higher minimum surface temperatures.

These outdoor experiments confirm that CPP-coated fiberglass structures offer superior passive radiative cooling performance under both controlled and natural outdoor conditions. The consistent temperature reduction of 5–15 °C compared to ambient, and favorable comparison to both inorganic fillers and commercial materials, highlight the practical viability of CPP as a high-performance, low-cost, and scalable cooling material for outdoor thermal management applications.

### 3.3. *Mechanical and Durability Test*

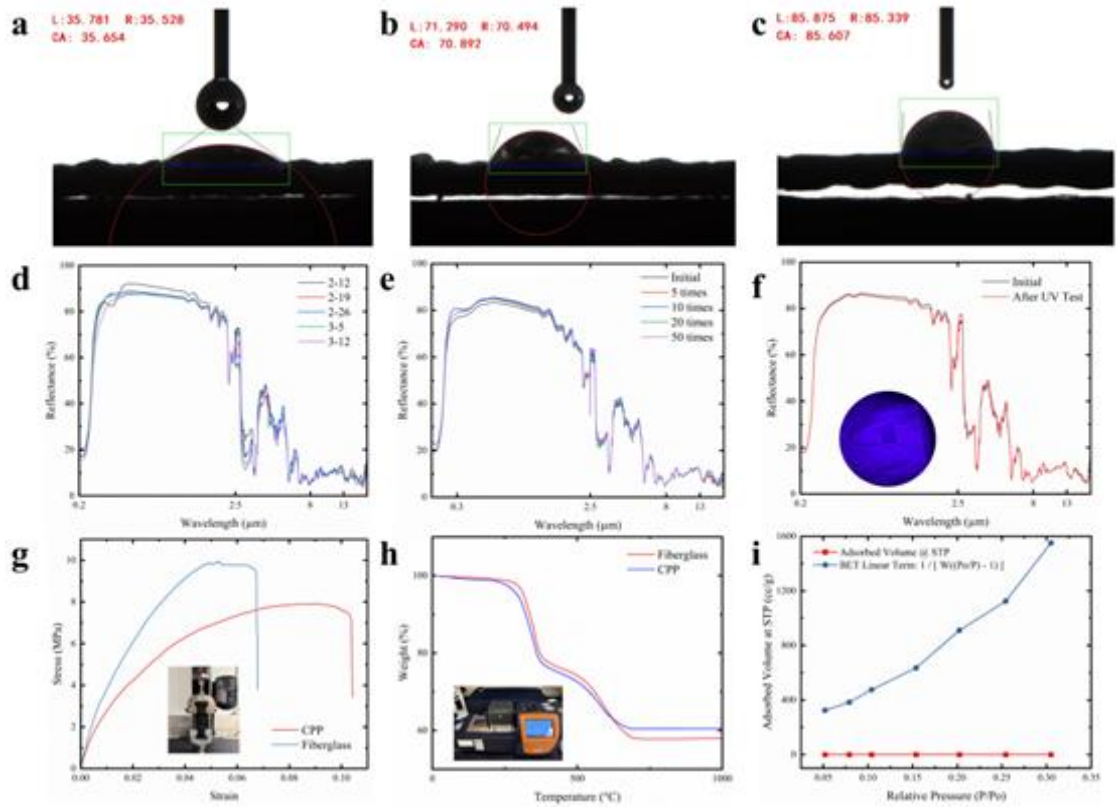

**Figure 4.** Durability, mechanical, and structural characterization of the CPP-coated fiberglass cast. (a–c) Water contact angles of PVA (~35°) and PMMA (~70°) layers, showing transition from hydrophilic to moderately hydrophobic behavior (~85°). (d) Four-week outdoor spectral durability test on the coated sample under varied weather conditions; spectra remain stable with minor decrease due to surface dust. (e) Abrasion test using 100 g weight on 1200-grid sandpaper for 10 m total distance; mid-IR emittance remains unchanged, solar reflectance slightly reduced. (f) UV durability test with SEVOU 365 nm 15 W UV light for 50 hours; negligible change in solar and mid-IR reflectance. (g) Tensile test results showing increased strain and reduced peak stress after CPP coating; coating adds flexibility while reducing stiffness. (h) TGA curves comparing coated and uncoated fiberglass casts; the coated sample exhibits enhanced thermal stability above 650 °C. (i) BET surface area analysis via $N_2$ adsorption at 77 K; linear fit yields a surface area of 0.740 m²/g, indicating a dense, low-porosity structure.

In practical applications, particularly in outdoor environments, water resistance and long-term material stability are essential for PDRC coatings. To assess the hydrophobicity of the bilayer structure, static water contact angle measurements were performed on both single-layer and double-layer samples. As shown in Figure 4a, the pristine fiberglass cast exhibited severe water permeability, with water droplets rapidly penetrating the porous matrix within seconds, rendering it unmeasurable by conventional contact angle methods. The single-layer PVA coating yielded a contact angle of approximately 35°, reflecting its strong hydrophilic nature, which is consistent with its water-based chemistry. In contrast, the PMMA-coated sample (Figure 4b) displayed a substantially higher contact angle of ~70°, attributed to its polymeric, nonpolar backbone that resists water adhesion.

To enhance moisture protection, CPP was incorporated into both layers, yielding a bilayer composite with a water contact angle of up to 85° (Figure S2). Contact angles above 80° are generally considered to indicate effective hydrophobic performance suitable for outdoor functional coatings, as supported by recent literature [26-27]. This improvement highlights the role of both PMMA and CPP in enhancing surface water resistance.

To evaluate environmental durability, a four-week outdoor exposure test was conducted on the Northeastern University campus (42.3384° N, 71.0890° W) from February 12, 2025, to March 12, 2025. The sample was subjected to a broad range of weather conditions, including sunshine, rainfall, snowfall, and strong winds. Weekly spectral reflectance measurements were collected to monitor coating stability. As illustrated in Figure 4c, reflectance spectra remained largely stable throughout the testing period. A slight initial drop was observed after the first week, which was attributed to dust accumulation on the surface. Importantly, no cleaning procedures were applied, allowing the sample to simulate real-world aging conditions.

Mechanical resistance was further evaluated through abrasion testing. A 100 g mass was applied to the surface of the coated sample, which was dragged across a 1200-grit sandpaper substrate for 20 cm per cycle. This procedure was repeated up to 50 cycles (equivalent to a total abrasion distance of 10 m). As shown in Figure 4d, the sample retained its mid-infrared emissivity throughout the test, and only minor reductions in solar reflectance were detected after extended abrasion, demonstrating good wear resistance.

Given the device's exposure to sunlight, UV stability was also assessed. A SEVOU 365 nm 15 W ultraviolet LED flashlight (Figure 4e) was positioned 30 cm above the sample, irradiating the surface continuously for 50 hours. Figure 4f presents the spectral results before and after

UV exposure. Minimal changes were observed, with the solar reflectance exhibiting a negligible decline, and mid-infrared performance remaining intact, demonstrating strong photo-stability under prolonged UV stress.

Mechanical deformation behavior was further characterized by using a universal tensile testing system (Instron 2710-103, Figure 4g). Both uncoated and CPP-coated fiberglass specimens were subjected to uniaxial tensile loading to determine their stress–strain response. The addition of the bilayer coating altered both ductility and stiffness. The engineering stress $\sigma$ and strain $\varepsilon$ were calculated using:

$$\sigma = \frac{F}{A} \tag{2}$$

$$\varepsilon = \frac{\Delta L}{L_0} \tag{3}$$

where $F$ is the tensile force, $A$ is the initial cross-sectional area, $\Delta L$ is the elongation, and $L_0$ is the original gauge length. In the linear elastic region, the relationship follows Hooke's Law:

$$\sigma = E \cdot \varepsilon \tag{4}$$

where $E$ is the Young's modulus. The coated specimens exhibited increased strain and reduced peak stress compared to uncoated samples, indicating enhanced flexibility at the expense of tensile strength. This trade-off is favorable for conformal surface applications where mechanical compliance is prioritized.

To evaluate structural reliability, the safety factor $n$ was computed as:

$$n = \frac{\sigma_{failure}}{\sigma_{working}} \tag{5}$$

where $\sigma_{failure}$ is the failure stress obtained from the test and $\sigma_{working}$ is the expected operational stress.

To assess potential rate-dependence of the mechanical response, a strain sensitivity model was considered:

$$\sigma = \sigma_0 \left(1 + \left(\frac{\dot{\varepsilon}}{C}\right)^{\frac{1}{P}}\right) \tag{6}$$

where $\sigma_0$ is the yield stress at quasi-static strain rate $\dot{\varepsilon}$, $C$ and $P$ are material-specific constants reflecting viscoelastic behavior. While constant-rate testing was employed in this study, this model provides a framework for future rate-dependent investigations.

Furthermore, the coating-substrate interface can be interpreted using the rule of mixtures for composite materials:

$$\sigma_{composite} = V_f \cdot \sigma_f + V_m \cdot \sigma_m \tag{7}$$

where $V_f$ and $V_m$ are the volume fraction, and $\sigma_f$ and $\sigma_m$ are the strengths of the filler (CPP coating) and matrix (fiberglass cast), respectively. This equation highlights the tunability of mechanical behavior via control of the component ratios.

Thermal stability was examined using TGA performed on a TA Instruments SDT650 system. The thermal decomposition profiles of both the uncoated and CPP-coated fiberglass casts were recorded under nitrogen. As shown in Figure 4h, both samples exhibited comparable thermal responses up to ~200 °C. In the intermediate range (200–650 °C), the uncoated sample retained marginally higher mass; however, beyond 650 °C, the CPP-coated sample demonstrated superior thermal stability, retaining more residual mass at elevated temperatures.

Surface area and porosity were further quantified by nitrogen adsorption–desorption analysis using a Quantachrome NOVA 2200e surface area analyzer (Figure 4i). The test was conducted at 77.35 K after 2 hours of degassing at 50 °C. BET analysis yielded a linear fitting slope of 4708.52 g$^{-1}$ and an intercept of −1.226 g$^{-1}$ over the relative pressure range of 0.05–0.30, resulting in a calculated BET surface area of 0.740 m²/g. The BET equation used for fitting is:

$$\frac{1}{W\left(\frac{P_0}{P} - 1\right)} = \frac{C - 1}{W_m C} \cdot \frac{P}{P_0} + \frac{1}{W_m C} \tag{8}$$

where $W$ is the volume of nitrogen adsorbed at a given relative pressure $P/P_0$, $W_m$ is the monolayer adsorbed gas quantity, and $C$ is the BET constant associated with the heat of adsorption. The high correlation coefficient ($R^2 = 0.9854$) confirmed the reliability of the fit. The low surface area is consistent with a dense, nonporous coating morphology that resists moisture uptake and contributes to structural integrity.

## 4. Conclusion

In this study, a bilayer PDRC coating was successfully integrated onto a commercially available fiberglass cast, forming a structurally robust, environmentally durable, and optically efficient cooling material. The dual-layer configuration, comprising a PVA-based adhesion layer and a PMMA-based protective layer, demonstrated excellent conformity to the irregular surface of the fiberglass substrate, as well as favorable water resistance, mechanical resilience, and solar reflectance when embedded with CPP particles.

Through comprehensive spectral analysis, the optimized 30 wt% CPP-loaded coating achieved over 90% solar reflectance in the 0.3–2.5 μm range and high mid-infrared emittance within the 8–13 μm atmospheric window, enabling substantial sub-ambient surface temperatures under outdoor exposure. In both wind-shielded and natural outdoor conditions, the coated samples maintained temperature differences of up to 15 °C below ambient, significantly outperforming reference materials including $TiO_2$, ZnO, and commercially painted wood.

Beyond thermal performance, the coating exhibited long-term stability against UV exposure, rainfall, and mechanical abrasion over a four-week outdoor durability test. Static water contact angle measurements confirmed a transition from hydrophilic to hydrophobic behavior with the addition of CPP and PMMA, reaching a peak angle of 85°, indicative of effective surface water repellency [28-29].

Mechanical evaluations revealed that the CPP coating enhanced surface flexibility and strain tolerance, while preserving sufficient tensile strength for practical applications. TGA analysis confirmed the composite's thermal robustness up to 700 °C, and BET testing verified a dense, low-porosity structure with a surface area of 0.740 m²/g.

Altogether, the results demonstrate that the proposed CPP-based PDRC-fiberglass composite offers a compelling solution for biocompatible, low-cost, scalable, and multifunctional thermal management in outdoor settings. Its integration-friendly processing, passive cooling performance, and environmental resilience render it a viable candidate for deployment in building envelopes, infrastructure coatings, and portable thermal shielding technologies.

**Acknowledgments**

This project is partially supported by the National Science Foundation through grant number CBET-1941743.

**Conflict of Interest**

The authors don't have any conflict of interest.

**Author Contributions**

Xuguang Zhang: Conceptualization, Data curation, Resources, Investigation, Methodology, Validation, Writing – original draft, Writing – review & editing. Hexiang Zhang: Resources, Writing – review & editing. Hanqing Liu: Resources, Writing – review & editing. Xiaoli Li: Resources, Writing – review & editing. Ying Mu: Resources, Writing – review & editing. Yutian Yang: Resources, Writing – review & editing. Marilyn L. Minus: Writing – review & editing. Ming Su: Writing – review & editing. Yi Zheng: Conceptualization, Resources, Funding acquisition, Writing – review & editing, Project administration, Supervision.